# Linking Quasi-Normal and Natural Modes of an open cavity

A. Settimi[(1)], S. Severini[(2)].

[(1)] Istituto Nazionale di Geofisica e Vulcanologia (INGV) –

via di Vigna Murata 605, 00143 Rome, Italy

[(2)] Centro Interforze Studi Applicazioni Militari (CISAM) –

Via della Bigattiera 10, 56122 San Piero a Grado, Pisa, Italy

# Abstract

The present paper proposes a comparison between the extinction theorem and the Sturm-Liouville theory approaches for calculating the e.m. field inside an optical cavity. We discuss for the first time to the best of our knowledge, in the framework of classical electrodynamics, a simple link between the Quasi Normal Modes (QNMs) and the Natural Modes (NMs) for one-dimensional (1D), two-sided, open cavities. The QNM eigenfrequencies and eigenfunctions are calculated for a linear Fabry-Perot (FP) cavity. The first-order Born approximation is applied to the same cavity in order to compare the first-order Born approximated and the actual QNM eigenfunctions of the cavity. We demonstrate that the first-order Born approximation for an FP cavity introduces symmetry breaking: in fact, each Born approximated QNM eigenfunction produces values below or above the actual QNM eigenfunction value on the terminal surfaces of the same cavity. Consequently, the two error-functions for an approximated QNM are not equal in proximity to the two terminal surfaces of the cavity.

**OCIS codes:**

260.2110 Electromagnetic optics

000.3860 Mathematical methods in physics

030.4070 Modes

260.5740 Resonance

050.2230 Fabry-Perot

050.5298 Photonic crystals

# 1. Introductory review.

The one dimensional (1D) theory of natural modes has been addressed through the application of at least two related approaches: the extinction theorem and the Sturm-Liouville theory.

- The extinction theorem implements the integro-differential equations of scattering and translates the problem into eigenvalues terms for the integro-differential equations. The principal difference from the classical theory of integral equations is that the eigenvalues bring nonlinearity into consideration.

- The discussion of mathematical methods for the classical Sturm Liouville theory of 1D problems generates a "Sturm-Liouville" theory for special 1D problems. The eigenfunctions can be demonstrated to be overcomplete, allowing the expansion of the field into natural modes inside, and at least up to the boundary of the open cavity.

**1.1. Extinction theorem approach (Natural Modes).**

A general definition of Natural Modes (NMs) of an electromagnetic (e.m.) or scalar field arising in relation to scattering issues has been given by Pattanayak and Wolf [1]. They defined NMs for a body of any given constitutive relations, bounded by a closed surface *S,* as the set of well-behaved outgoing solutions of Maxwell equations that obey the continuity conditions at the boundary *S* (a similar definition can be formulated for the NMs of potential scattering). The solutions of Maxwell equations satisfying these conditions were demonstrated by Wolf and Pattanayak [2] to be the eigenfunctions of the homogeneous portion of the integro-differential equations of e.m. scattering with suitable eigenvalues, or solutions to Maxwell equations that satisfy a nonlocal boundary condition. These two formulations are considered equivalent.

This definition is a generalization of previous proposals [3-6] which were all negatively influenced by the application of a "preconceived" notion of the structure of the resonance states [7]. This problem was avoided in the work of Pattanayak and Wolf who were able to achieve a definition for natural modes without any "preconceived" notions and directly from the scattering integral equations.

A further important general property of NMs is that they can exist in a medium without the presence of a driving field. This can be explained in terms of internal (electron) oscillations of the scatterer: the oscillations continue for some time even after the driving field is gone.

Theoretical results were obtained by Hoenders in refs. [8-11], developing a Hilbert-Schmidt type theory, leading to a bilinear expansion of the NMs of the resolvent kernel connected with the integral equations of e.m. and potential scattering theory (see references therein [8-11]). The initial value problem addressed in refs. [8-11] could be solved using this Hilbert-Schmidt type of theory, implementing the temporal Laplace transform of the Maxwell equations.

Theoretical and applicational results were provided by Hoenders and Bertolotti in ref. [12] developing a scattering theory for finite 1D-PBG structures [13] in terms of the NMs of the scatterer. This theory generalizes the classical Hilbert-Schmidt type bilinear expansions of the field propagator to a bilinear expansion into the NMs (see references therein [12]). It is shown that the Sturm-Liouville type of expansions for dispersive media differ considerably from those for non-dispersive media, for example they are overcomplete.

**1.2. Sturm-Liouville theory approach (Quasi Normal Modes).**

Several authors have discussed the issue of field description inside an open cavity. Leung et al. in refs. [14-20] proposed the description of the e.m. field, in one-sided open microcavities, in terms of Quasi Normal Modes (QNMs). Microcavities are mesoscopic open systems, which means that they are "leaky", or non-conservative, and consequently the eigenfunctions of the resonance field are QNMs with complex eigenfrequencies. These significantly influence numerous optical processes and the analogy with normal modes of conservative systems is marked. In some cases, the QNMs form a complete set and much of the standard formalism can be implemented. Microcavities are also open systems that require some output coupling. As a consequence of the output coupling, the e.m. field and energy in the microcavity itself would be continuously lost to the outside. Therefore, the microcavity is a non conservative system in physical terms, and, from a mathematical perspective, the operators which emerge would be non-Hermitian (or not self-adjoint). This generates the interesting challenge of generalizing the familiar tools of mathematical physics to setting of such a non Hermitian nature. The issues that emerge, and the framework developed to address them, are generic to many other open system contexts.

In refs. [21,22], the QNM approach was applied and extended to a description of scalar field behaviour in double-sided open optical cavities, in particular 1D-PBG structures. The validity of the approach is discussed by proving QNM completeness, considering the distribution of complex frequencies and the corresponding field distributions.

In ref. [23], the e.m. field inside an optical open cavity was analyzed within the framework of the QNM theory and the role of complex frequencies in the transmission coefficient is clarified. Application to a quarter-wave (QW) symmetric 1D-PBG structure is discussed to illustrate the usefulness and meaningfulness of the results.

Finally, in ref. [24], demonstrations are provided for a large number of theoretical assumptions on QNM metrics for an open cavity, and a simple, direct method for calculating the QNM norm for a 1D-PBG structure is reported.

Applicational results were provided by Bertolotti in ref. [25], discussing the linear properties of one dimensional resonant cavities and using matrix and ray methods (see references therein [25]). Fabry-Perot (FP), Photonic Crystals and 1D cavities in PBG structures are considered. The QNMs for the description of the e.m. field in open cavities are introduced and some applications are reported.

## 1.3. Topic and structure of the paper.

The present paper proposes a comparison between the extinction theorem [1,2][8-11] and the Sturm-Liouville theory [14-20][21-24] approaches for calculating the e.m. field inside an optical cavity We discuss for the first time to the best of our knowledge, in the framework of classical electrodynamics, a simple link between the Quasi Normal Modes (QNMs) and the Natural Modes (NMs) for one-dimensional (1D), two-sided, open cavities. The QNM eigenfrequencies and eigenfunctions are calculated for a linear Fabry-Perot (FP) cavity. The first-order Born approximation is applied to the same cavity in order to compare the first-order Born approximated and the actual QNM eigenfunctions of the cavity. We demonstrate that the first-order Born approximation for an FP cavity introduces symmetry breaking: in fact, each Born approximated QNM eigenfunction produces values below or above the actual QNM eigenfunction value on the terminal surfaces of the same cavity Consequently, the two error-functions for an approximated QNM are not equal in proximity to the two terminal surfaces of the cavity.

The paper is organized as follows. Section 2 reiterates the definition for QNMs in an open cavity without external pumping. The QNMs occur as poles of the transformed Green function with negative imaginary parts. In section 3, the QNM approach is applied to a linear FP cavity. Sections 4 and 5 are dedicated to a critical presentation of the NMs for an open cavity as well as pointing out some useful mathematical and physical aspects of this theory. Section 6 describes the application of the first-order Born approximation to a linear FP cavity. Conclusions are drawn in section 7. Finally, an outline is presented of the somewhat lengthy calculations in the appendices.

## 2. The Quasi Normal Mode (QNM) approach.

In ref. [24], we formalized suitable "outgoing wave conditions" for an open cavity without external pumping. More explicitly than Leung's refs. [14-20], a Laplace transform of the e.m. field is considered, to take the cavity leakages into account, and it is noted that, only in the complex domain defined by a $\pi/2$-rotated Laplace transform [26], can the Quasi Normal Modes (QNMs) be defined as the poles of the transformed Green function, with negative imaginary part. Moreover, ref. [24] clarified the role of the complex Green function formalism for the definition of the QNMs; these considerations are reported there in more detail compared to those already presented in refs. [21,22][23].

With reference to fig. 1, an open cavity of length $d$ is considered, filled with a material of refractive index $n(x)$ and enclosed in an infinite external space. The cavity also includes the terminal surfaces, so it is represented as $C = [0,d]$ and the rest of universe as $U = (-\infty, 0) \cup (d, \infty)$. The refractive index is characterized by [14-20]:

- the *discontinuity conditions*, i.e. $n(x)$ presents a step at $x=0$ and $x=d$ (in this hypothesis a natural demarcation of a finite region is provided);

- the *no tail conditions*, i.e. $n(x) = n_0$ for $x < 0$ and $x > d$ (in this hypothesis outgoing waves are not scattered back).

The electromagnetic (e.m.) field $E(x,t)$ in the open cavity solves the equation [27]

$$\left[\frac{\partial^2}{\partial x^2} - \rho(x)\frac{\partial^2}{\partial t^2}\right] E(x,t) = 0, \qquad (2.1)$$

where $\rho(x) = [n(x)/c]^2$, and $c$ is the speed of light in vacuum. If there is no external pumping, the e.m. field satisfies suitable "outgoing waves conditions" [21-24] (see fig. 1)

$$\partial_x E(x,t) = \sqrt{\rho_0}\partial_t E(x,t) \quad \text{for} \quad x < 0, \qquad (2.2)$$

$$\partial_x E(x,t) = -\sqrt{\rho_0}\partial_t E(x,t) \quad \text{for} \quad x > d, \qquad (2.3)$$

where $\rho_0 = (n_0/c)^2$, and $n_0$ is the outside refractive index. In fact, on the left side of the same cavity, i.e. $x < 0$, the e.m. field is travelling in the negative sense of the *x*-axis, i.e. $E(x,t) = E[x + (c/n_0)t]$ so eq. (2.2) holds as one can easily demonstrate. On the right side of the cavity, i.e. $x > d$, the e.m. field is travelling in the positive sense of the *x*-axis, i.e. $E(x,t) = E[x - (c/n_0)t]$ so eq. (2.3) holds as well.

The *Quasi Normal Modes* pairs $[\omega_n, f_n(x)]$ are specified by:

- complex eigenfrequencies $\omega_n$, $n \in Z = \{0, \pm 1, \pm 2, ...\}$, with negative imaginary parts $\text{Im}\,\omega_n < 0$, so they are the *non-stationary modes* of an open cavity [14-20] (they are observed in the frequency domain as resonances of finite width or in the time domain as damped oscillations);

- and, under the "discontinuity and no tail conditions", by eigenfunctions $f_n(x)$ which form an *orthogonal basis only inside the open cavity* [14-20] (it is possible to describe the QNMs in a manner parallel to the normal modes of a closed cavity, with respect a definition of the norm introduced later in the text) [see eq. (3.10)].

It follows that the QNMs $[\omega_n, f_n(x)]$ solve the equation [27]:

$$\left[\frac{d^2}{dx^2} + \omega_n^2 \rho(x)\right] f_n(x) = 0. \tag{2.4}$$

Moreover, only for long distances from the open cavity, the QNMs do not represent e.m. fields because they satisfy the QNM "asymptotic conditions" [21-24],

$$f_n(x) = \exp(\pm i\omega_n \sqrt{\rho_0} x) \to \infty \quad \text{for} \quad x \to \pm\infty, \tag{2.5}$$

while, inside the same cavity and near its terminal surfaces from outside, the QNMs represent non-stationary modes, in fact the asymptotic conditions (2.5) imply the "formal" QNM "outgoing waves" conditions [21-24]:

$$\partial_x f_n(x)\big|_{x=0^-} = -i\omega_n \sqrt{\rho_0}\, f_n(0^-), \tag{2.6}$$

$$\partial_x f_n(x)\big|_{x=d^+} = i\omega_n \sqrt{\rho_0}\, f_n(d^+). \tag{2.7}$$

The conditions (2.6) and (2.7) are called "formal" because they refer to the QNMs which do not represent e.m. fields for long distances from the open cavity, and "outgoing waves" because they are formally identical to the real outgoing waves conditions for the e.m. field next to the surfaces of the cavity. In fact, eqs. (2.6) and (2.7) for the QNMs can be derived if and only if the requirement of outgoing waves holds for the e.m. field.

## 3. QNM eigenfrequencies and eigenfunctions of a linear Fabry-Perot cavity.

In this section, the Quasi Normal Mode (QNM) approach is applied to a linear Fabry-Perot (FP) cavity.

Suppose that the refractive index of the material inside the same linear cavity is [27]:

$$n(x) = \begin{cases} n , & x \in [0,d] \\ n_0 , & \text{outside} \end{cases}. \tag{3.1}$$

If the e.m. field equation (2.4) is solved, adding the outgoing waves conditions (2.6) and (2.7), the QNM eigenfrequencies of the FP cavity are calculated, as shown in the following relations

$$\omega_k = k\alpha - i\beta, \tag{3.2}$$

where

$$\alpha = \frac{\pi}{\sqrt{\rho}d} , \quad \beta = \frac{1}{\sqrt{\rho}d}\ln\left(\frac{n+n_0}{n-n_0}\right) , \quad k \in Z = \{0, \pm 1, \pm 2, ...\}, \tag{3.3}$$

with $\rho = (n/c)^2$ and $c$ the speed of light in vacuum.

The QNM eigenfunctions of the same cavity are in turn expressible as

$$f_k(x) = N \cdot \left[\cos\left(\omega_k \sqrt{\rho}x\right) - i\frac{n_0}{n}\sin\left(\omega_k \sqrt{\rho}x\right)\right], \tag{3.4}$$

where $N$ is a suitable constant of normalization.

Some useful results are reported, involving the QNM frequencies (3.2)-(3.3):

$$\cos\left(\omega_k \sqrt{\rho}d\right) = (-1)^k \frac{\rho+\rho_0}{\Delta\rho} , \quad \sin\left(\omega_k \sqrt{\rho}d\right) = -i(-1)^k \frac{2\sqrt{\rho \cdot \rho_0}}{\Delta\rho}, \tag{3.5}$$

where $\Delta\rho = \rho - \rho_0$, with $\rho = (n/c)^2$ and $\rho_0 = (n_0/c)^2$.

Applying eqs. (3.5), in the case of a linear FP cavity, it is easy to demonstrate that the QNM frequencies (3.2) and the QNM functions (3.4) of the same linear cavity satisfy the general properties [14-20]

$$\begin{aligned} \omega_{-k} &= -\omega_k^* \\ f_{-k}(x) &= f_k^*(x) \end{aligned} ; \qquad (3.6)$$

moreover, the outgoing wave conditions (2.6) and (2.7), referred to the QNMs, can be cast in symmetric relations, as follows:

$$f_k(d) = (-1)^k f_k(0), \qquad (3.7)$$

$$\partial_x f_k(x)\big|_{x=d} = -(-1)^k \partial_x f_k(x)\big|_{x=0}. \qquad (3.8)$$

A useful integral is $I_{p,q} = \int_0^d \rho(x) f_p(x) f_q(x) dx$ which can be calculated as

$$I_{p,q} = N^2 \cdot \rho \left[ \delta_{p,q} \frac{\Delta \rho}{2\rho} d + \frac{n_0}{n} \frac{1+(-1)^{p+q}}{i(\omega_p + \omega_q)\sqrt{\rho}} \right], \qquad (3.9)$$

where $\delta_{p,q}$ is the Kronecker delta. Eq. (3.9) is derived after some algebra, first inserting eq. (3.4) and then applying eqs. (3.5).

The norm of the QNM function (3.4) is a complex number, given by the following expression [14-20][21-24]

$$\langle f_k | f_k \rangle = 2\omega_k \int_0^d \rho(x) f_k^2(x) dx + i\sqrt{\rho_0} [f_k^2(0) + f_k^2(d)], \qquad (3.10)$$

the main difference with the ordinary definition of the norm being the presence of $f_k^2(x)$ rather than $|f_k(x)|^2$ and the two additive "surfaces terms" $i\sqrt{\rho_0} f_k^2(0)$ and $i\sqrt{\rho_0} f_k^2(d)$.

Applying eq. (3.9), in the case of a linear FP cavity, it is easy to demonstrate that the QNM norm (3.10) for the same cavity with QNM frequencies (3.2)-(3.3) and QNM functions (3.4) assumes the form:

$$\langle f_k | f_k \rangle = N^2 \cdot \omega_k \Delta \rho d. \qquad (3.11)$$

If a normalized version of the QNM function is adopted, corresponding to $\langle f_k | f_k \rangle = 2\omega_k$, then the constant $N$ is fixed, $N^2 = \dfrac{2}{\Delta\rho \cdot d}$, so that:

$$f_k(0) = N = \sqrt{\dfrac{2}{\Delta\rho \cdot d}} \quad , \quad f_k(d) = (-1)^k N \ . \qquad (3.12)$$

## 4. A deeper examination of the Natural Mode (NM) approach.

In comparison with the choice of Pattanayak and Wolf in refs. [1,2] that the Green function of the 3D universe satisfies some local "outgoing wave conditions" at infinity, in this section we underline that it is possible to define a suitable equivalent Feynman complex propagator of the one-dimensional (1D) universe which satisfies some other local "incoming wave conditions" at infinity. Then, unlike the Hoenders and Bertolotti assertion in ref. [12] that the class of Natural Modes (NMs) [8-11] for an open cavity should coincide with the Quasi Natural Modes (QNMs) [25] of the cavity, we conjecture that, even if the NMs do not represent electromagnetic (e.m.) fields for long distances from the open cavity, in any case they can be obtained only in the context of the incoming waves hypothesis for the e.m. field.

Finally, some remarkable relations are demonstrated, permitting evaluation of the NM eigenfunctions of an open cavity next to the surfaces of the same cavity.

### 4.1. Green function of the one dimensional (1D) universe.

The one dimensional (1D) universe is modelled as an infinite 1D cavity, $C_\infty = \{x | x \in (-L/2, L/2) \ , \ L \to \infty\}$, which is filled by a linear, isotropic and homogeneous dielectric, with a dielectric constant $\varepsilon = \varepsilon_0 \varepsilon_r$ and a magnetic permeability $\mu = \mu_0 \mu_r \cong \mu_0$.

The Green function of the 1D universe [27] solves the equation

$$\left(\frac{\partial^2}{\partial x^2}+\rho_0\omega^2\right)G_0(x,x',\omega)=-\delta(x-x'), \tag{4.1}$$

and satisfies the symmetry property [27]:

$$G_0(x,x',\omega)=G_0(x',x,\omega). \tag{4.2}$$

In the Green function (4.1), the e.m. field $G_0(x,x',\omega)$ at the observation point $x$, when a Dirac delta source is allocated in the $x'$ point, presents the same form as the e.m. field $G_0(x',x,\omega)$ when $x$ and $x'$ are inverted, i.e. eq. (4.2) holds.

Adopting the Feynman prescription [28], the Green function $G_0(x,x',\omega)$ which solves the eq. (4.1) and satisfies the symmetry property (4.2) can be replaced by the complex propagator

$$G_0^{(F)}(x,x',\omega)=-\frac{i}{2\omega\sqrt{\rho_0}}\left\{\exp[-i\omega\sqrt{\rho_0}(x-x')]\theta(x-x')+\exp[-i\omega\sqrt{\rho_0}(x'-x)]\theta(x'-x)\right\}, \tag{4.3}$$

where the unit step function $\theta(x)$ is defined as: $\theta(x)=\begin{cases}0, & x<0 \\ 1, & x\geq 1\end{cases}$.

The Feynman complex propagator $G_0^{(F)}(x,x',\omega)$ expressed by eq. (4.3) satisfies the boundary conditions

$$\partial_x G_0^{(F)}(x,x',\omega)=i\omega\sqrt{\rho_0}G_0^{(F)}(x,x',\omega) \quad , \quad \forall x'\leq x, \tag{4.4}$$

$$\partial_x G_0^{(F)}(x,x',\omega)=-i\omega\sqrt{\rho_0}G_0^{(F)}(x,x',\omega) \quad , \quad \forall x'\geq x, \tag{4.5}$$

where $\rho_0=(n_0/c)^2$, with $n_0\cong\sqrt{\varepsilon_r}$ the refractive index of the dielectric material and $c=1/\sqrt{\mu_0\varepsilon_0}$ the speed of light in vacuum.

In the Feynman propagator (4.3), if the observation point is allocated to $x$, then the e.m. field $G_0^{(F)}(x,x',\omega)$ is incoming near the point $x$, i.e. it is travelling with a velocity component $1/\sqrt{\rho_0}$ on the left side $x'\leq x$ [eq. (4.4) holds], and with a velocity component $(-1/\sqrt{\rho_0})$ on the right side $x'\geq x$ [eq. (4.5) holds]. In other words, the propagator $G_0^{(F)}(x,x',\omega)$ can be interpreted as an e.m. field which is scattering near an infinitesimal range of the point $x$ (Appendix A).

**4.2. NM definition.**

Start again from the same cavity of length $d$, filled with a refractive index $n(x)$ and enclosed in an infinite external space [fig.1].

The Natural Modes (NMs) of the open cavity [1,2] are defined as some eigensolutions $[\omega'_n, \psi_n^{(nat)}(x)]$ with $n \in Z = \{0, \pm 1, \pm 2, \ldots\}$ such that their eigenfrequencies $\omega'_n$ are complex, i.e. $\omega'_n \in C$, and their eigenfunctions $\psi_n^{(nat)}(x)$ solve the homogeneous integral equation inside the cavity [8-11][12]

$$\psi_n^{(nat)}(x) = \int_{0^-}^{d^+} (\omega'_n)^2 [\rho(x') - \rho_0] \psi_n^{(nat)}(x') G_0^{(F)}(x, x', \omega'_n) dx' \quad, \quad 0 \le x \le d \;, \tag{4.6}$$

where $\rho(x) = [n(x)/c]^2$.

Eq. (4.6) is the basic equation for determining the NM eigenfunctions and is characterized by a form which can be deduced from the integral equation of potential scattering [27]. It should be noted, however, that it is only an integral equation for the NM eigenfunctions at the points within the cavity, which, in the context of eq. (4.6), can be described as a scattering volume. It seems therefore reasonable to think that any solution of eq. (4.6) should satisfy the continuity conditions with suitable e.m. fields that exhibit incoming wave behaviour near the open cavity, interpreted as a scatterer.

*4.2.a. First theorem on NMs.*

Next, the extinction theorem (NMs) and the Sturm-Liouville theory (QNMs) approaches are compared. The class of NMs for an open cavity does not coincide with the QNMs of the same cavity, in the sense that the NMs $[\omega'_n, \psi_n^{(nat)}(x)]$, with $n \in Z = \{0, \pm 1, \pm 2, \ldots\}$, defined by eq. (4.6), satisfy the e.m. problem consisting of the differential Helmholtz equation inside the open cavity [which is similar to eq. (2.4)] [8-11],

$$\left[\frac{\partial^2}{\partial x^2} + (\omega'_n)^2 \rho(x)\right] \psi_n^{(nat)}(x) = 0 \quad , \quad 0 \leq x \leq d , \qquad (4.7)$$

but with the boundary conditions next to the terminal surfaces of the cavity [which are not the boundary conditions of eqs. (2.6) and (2.7)]:

$$\partial_x \psi_n^{(nat)}(x)\big|_{x=0^-} = i\omega'_n \sqrt{\rho_0} \psi_n^{(nat)}(0^-) , \qquad (4.8)$$

$$\partial_x \psi_n^{(nat)}(x)\big|_{x=d^+} = -i\omega'_n \sqrt{\rho_0} \psi_n^{(nat)}(d^+) . \qquad (4.9)$$

The demonstration of these assertions is given in Appendix B.

A more refined mathematical analysis shows that the expressions on the right side of eqs. (4.8) and (4.9) are "formally" of opposite sign respectively with eqs. (2.6) and (2.7), so it is natural to characterize eqs. (4.8) and (4.9) as the "formal" NM "incoming waves" conditions if eqs. (2.6) and (2.7) have been defined as the "formal" QNM "outgoing waves" conditions. According to eqs. (4.8) and (4.9), it can be postulated that, even if the NMs do not represent e.m. fields for long distance from the open cavity, in any case they may be obtained only in the context of the incoming waves hypothesis for the e.m. field. This hypothesis should confirm the above hypothesis according to which any solution of the integral equation (4.6) satisfies the continuity conditions with suitable e.m. fields that exhibit an incoming wave behaviour near the cavity, interpreted as a scatterer.

*4.2.b. Second theorem on NMs.*

Some remarkable relations permit evaluation of the NM eigenfunctions of an open cavity next to the surfaces of the same cavity:

$$\psi_n^{(nat)}(0^-) = \frac{1}{2i\omega'_n \sqrt{\rho_0}} \int_{0^-}^{d^+} (\omega'_n)^2 [\rho(x) - \rho_0] \psi_n^{(nat)}(x) \exp(-i\omega'_n \sqrt{\rho_0} x) dx , \qquad (4.10)$$

$$\psi_n^{(nat)}(d^+) = \frac{\exp(-i\omega_n'\sqrt{\rho_0}d)}{2i\omega_n'\sqrt{\rho_0}} \int_{0^-}^{d^+} (\omega_n')^2[\rho(x)-\rho_0]\psi_n^{(nat)}(x)\exp(i\omega_n'\sqrt{\rho_0}x)dx. \qquad (4.11)$$

The demonstration of these assertions is given in Appendix B.

## 5. Linking the NMs and the QNMs of an open cavity.

In this section, we describe and provide the proof and physical interpretation of a fundamental theorem on the Natural Modes (NMs). Going beyond refs. [21-24], we deduce that, if the open cavity is a medium which scatters very weakly, so that the refractive index $\rho(x)$ of the same cavity differs only slightly from the value $\rho_0$ in the universe, then the electromagnetic (e.m.) field which satisfies the "outgoing waves conditions" can be transformed in an e.m. field which satisfies some "incoming waves conditions" by simply changing the sign of the frequency. Unlike Hoenders and Bertolotti's ref. [12], a physical interpretation is provided for the link involving the NMs [8-11] and Quasi Normal Modes (QNMs) [25].

A further important general property of NMs is that they can exist in a medium without the presence of a driving field. This can be explained in terms of internal (electron) oscillations of the scatterer: the oscillations continue for some time even after the driving field is gone.

In addition, some remarkable relations regarding the NMs are converted in the dual relations for the QNMs.

### 5.1. A fundamental theorem on NMs.

If $[\omega_n', \psi_n^{(nat)}(x)]$ with $n \in Z = \{0, \pm 1, \pm 2, ...\}$ describe NMs [1,2] of an open cavity and are such that their eigenfunctions $\psi_n^{(nat)}(x)$ solve the homogeneous integral equation (4.6) inside the same cavity [8-11][12], then their eigenfrequencies $\omega'_n$ are characterized by positive imaginary parts:

$$\operatorname{Im}\omega'_n > 0. \tag{5.1}$$

The demonstration of this assertion is provided in Appendix C.

Below is a discussion of the interesting physical interpretation of the theorem just introduced. Again with reference to fig.1, if there are two external counter-propagating scattering fields, it can be supposed that, outside the open cavity, the e.m. field satisfies suitable incoming waves conditions [21-24]:

$$\partial_x E^*(x,t) = -\sqrt{\rho_0}\partial_t E^*(x,t) \quad for \quad x < 0. \tag{5.2}$$

$$\partial_x E^*(x,t) = \sqrt{\rho_0}\partial_t E^*(x,t) \quad for \quad x > d. \tag{5.3}$$

The incoming waves conditions (5.2) and (5.3) outside the cavity hold when the e.m. field

$$E^*(x,t) = E^*_P(x,t) + E^*_R(x,t) \quad , \quad \text{for } x < 0 \text{ and } x > d, \tag{5.4}$$

consisting of the external scattering field $E^*_P(x,t)$ and the reflected field $E^*_R(x,t)$, can be approximated only by the scattering field,

$$E^*(x,t) \cong E^*_P(x,t) \quad , \quad \text{for } x < 0 \text{ and } x > d, \tag{5.5}$$

because the reflected field is not very relevant,

$$E^*_R(x,t) \ll E^*_P(x,t) \quad , \quad \text{for } x < 0 \text{ and } x > d. \tag{5.6}$$

Under these circumstances it is plausible to assume that the open cavity is a medium which scatters very weakly [27], so that the refractive index $\rho(x)$ of the same cavity differs only slightly from the value $\rho_0$ in the universe.

If the $\pi/2$-rotated Laplace transform [24][26] is applied to the incoming waves conditions (5.2) and (5.3), it follows that:

$$\partial_x \tilde{E}^*(x,\omega) = i\omega\sqrt{\rho_0}\tilde{E}^*(x,\omega) \quad for \quad x < 0, \tag{5.7}$$

$$\partial_x \tilde{E}^*(x,\omega) = -i\omega\sqrt{\rho_0}\tilde{E}^*(x,\omega) \quad \text{for} \quad x > d. \tag{5.8}$$

The e.m. field $\tilde{E}^*(x,\omega)$ which satisfies the incoming waves conditions (5.7) and (5.8) can be transformed in the e.m. field $\tilde{E}(x,\omega)$ which satisfies some outgoing waves conditions by simply changing the sign of the frequency, i.e.

$$\tilde{E}^*(x,\omega) = \tilde{E}(x,-\omega), \tag{5.9}$$

because the incoming waves equations (5.7) and (5.8) can be converted into outgoing waves equations simply applying eq. (5.9). Thus, eq. (5.9) defines an application for which, each complex frequency $\omega$ is simply changed into the frequency of opposite sign $(-\omega)$.

Applying eq. (5.9), it follows that the NMs $[\omega'_n, \psi_n^{(nat)}(x)]$ and the QNMs $[\omega_n, f_n(x) = f(x, \omega_n)]$ are correlated by the links:

$$\omega'_n = -\omega_n, \tag{5.10}$$

$$\psi_n^{(nat)}(x) = f(x, -\omega'_n). \tag{5.11}$$

So, only for long distances from the open cavity, the NMs do not represent e.m. fields because they satisfy the NM "asymptotic conditions":

$$\psi_n^{(nat)}(x) = \exp(\mp i\omega'_n \sqrt{\rho_0} x) \to \infty \quad \text{for} \quad x \to \pm\infty. \tag{5.12}$$

Inside the same cavity and near its terminal surfaces from outside, the NMs do not represent stationary modes. In fact, the NM asymptotic conditions (5.12) imply the "formal" NM "incoming waves" conditions [see eqs. (4.8) and (4.9)]. The conditions (4.8) and (4.9) were referred to above as "formal" because they are referred to the NMs which, in accordance to the eq. (5.12), do not represent e.m. fields for long distances from the open cavity. The NMs can be seen as "incoming waves" because they are formally identical to the real incoming waves conditions (5.7), here defined for the e.m. field outside the cavity. In fact, the eqs. (4.8) and (4.9) for the NMs can be derived if and only if the requirement of incoming waves holds for the e.m. field.

Thus, following the physical interpretation of the fundamental theorem on NMs: the QNMs are defined for the problem of an e.m. field outgoing from an open cavity with a generic refractive index. It is natural to introduce the NMs only for the scattering problems of an e.m. field incoming within a medium that scatters rather weakly, i.e. when the refractive index of the medium differs only slightly from the value in the universe. By substituting $\omega'_n = -\omega_n$, the formal incoming waves conditions (4.8) and (4.9) for the NMs with eigenfrequencies $\omega'_n$ are reduced to the formal incoming waves conditions (2.6) and (2.7) for the QNMs with eigenfrequencies $\omega_n$. Moreover, the remarkable relations (4.10), (4.11) for the NMs can be converted in the dual relations for the QNMs:

$$f_n(0^-) = -\frac{1}{2i\omega_n\sqrt{\rho_0}} \int_{0^-}^{d^+} (\omega_n)^2 [\rho(x) - \rho_0] f_n(x) \exp(i\omega_n\sqrt{\rho_0} x) dx, \qquad (5.13)$$

$$f_n(d^+) = -\frac{\exp(i\omega_n\sqrt{\rho_0} d)}{2i\omega_n\sqrt{\rho_0}} \int_{0^-}^{d^+} (\omega_n)^2 [\rho(x) - \rho_0] f_n(x) \exp(-i\omega_n\sqrt{\rho_0} x) dx. \qquad (5.14)$$

Finally, if, outside the open cavity, there are two external counter-propagating pumping fields which satisfy the incoming waves conditions (5.2) and (5.3), inside the same cavity, they lead to two internal counter-propagating scattering fields which still satisfy some incoming waves conditions similar to eqs. (5.2) and (5.3). Even if, starting from a certain time, the two external counter-propagating pumping fields are suppressed, the two internal counter-propagating scattering fields continue to travel, obeying the incoming waves conditions; i.e. the system is not Markovian: the kernel of eq. (4.6) conserves a memory, so the conditions (4.8) and (4.9) have to persist. This section confirms the conjecture, advanced below eq. (4.6), for which any solution of the integral equation (4.6) satisfies the continuity conditions with suitable e.m fields that exhibit an incoming wave behaviour near the open cavity, interpreted as a scatterer; and the conjecture, advanced below eqs. (4.8) and (4.9), for which even if the NMs do not represent e.m. fields for outside the cavity, in any case they can be obtained only in the context of the incoming waves hypothesis for the e.m. field.

## 6. First-order Born approximation.

In this section, we apply the first-order Born approximation to a linear Fabry-Perot (FP) cavity proving that, compared to the approximation of the universe mode eigenfunctions [29], the first-order Born approximation for an FP cavity [27] better approximates the envelope of the Natural Mode (NM) eigenfunctions. In contrast with the universe mode approximation, the first-order Born approximation satisfies the "formal" NM "incoming waves" conditions.

From expressions (4.6) and (4.7) it is clear that if

$$\rho(x) = \rho_0 + \Delta\rho(x), \tag{6.1}$$

with

$$\Delta\rho(x) \begin{cases} \ll \rho_0 & \text{for } 0 \leq x \leq d \\ = 0 & \text{for } x < 0 \text{ and } x > d \end{cases}, \tag{6.2}$$

it is plausible to assume that a good approximation will be obtained for the NM eigenfunction $\psi_k^{(nat)}(x)$ on the left side of eq. (4.6) if the NM eigenfunction $\psi_k^{(nat)}(x')$ under the integral on the right side of eq. (4.6) is replaced by the eigenfunction for a mode of the universe [29]:

$$\begin{aligned} \psi_k^{(nat)}(x') &\simeq g_\omega(x')\big|_{\omega=\omega'_k} \\ g_\omega(x') &= \frac{1}{\sqrt{2\pi}} \exp(i\omega\sqrt{\rho_0}\, x') \end{aligned}. \tag{6.3}$$

A first approximation to the solution of the integral equation (4.6) is then obtained with the expression:

$$\psi_k^{(nat)}(x) \cong \int_{0^-}^{d^+} (\omega'_k)^2 \Delta\rho(x')\, g_\omega(x')\big|_{\omega=\omega'_k} G_0^{(F)}(x,x',\omega'_k)\, dx' \quad, \quad 0 \leq x \leq d. \tag{6.4}$$

Inserting, in the expression (6.4) for the NM eigenfunctions of the cavity, the expression (6.3) for the modes of the universe and the expression (4.3) for the Feynman propagator, eq. (6.4) can be explicited as

$$\psi_k^{(nat)}(x) \cong -\frac{i}{2\sqrt{2\pi}}\frac{\omega_k'}{\sqrt{\rho_0}}[A(x)\exp(i\omega_k'\sqrt{\rho_0}x) + B_k(x)\exp(-i\omega_k'\sqrt{\rho_0}x)] \ , \quad 0 \le x \le d, \quad (6.5)$$

with

$$A(x) = \int_0^d \Delta\rho(x')\theta(x'-x)dx', \quad (6.6)$$

$$B_k(x) = \int_0^d \exp(i2\omega_k'\sqrt{\rho_0}x')\Delta\rho(x')\theta(x-x')dx' . \quad (6.7)$$

### *6.1. Linear Fabry-Perot (FP) cavity.*

A Fabry-Perot (FP) cavity is now modelled as a finite cavity $C = [0,d]$ with a relative refractive index $n^2(x) \propto \rho(x)$ which is constant, i.e. [27]

$$\Delta\rho(x) = \Delta\rho = const, \quad 0 \le x \le d, \quad (6.8)$$

and it is supposed that the relative refractive index inside the Fabry-Perot is close to the value outside the cavity, i.e. [27]

$$\Delta\rho \ll \rho_0 . \quad (6.9)$$

Applying eqs. (6.8) and (6.9), the coefficients (6.6) and (6.7) can be calculated as

$$A(x) \stackrel{0 \le x \le d}{=} \Delta\rho(d-x) \quad (6.10)$$

and

$$B_k(x) \stackrel{0 \le x \le d}{=} \Delta\rho \frac{\exp(i2\omega_k'\sqrt{\rho_0}x) - 1}{i2\omega_k'\sqrt{\rho_0}}, \quad (6.11)$$

so the first-order Born approximation for the NM eigenfunctions (6.5) can be specified for a FP cavity as:

$$\psi_k^{(nat)}(x) \cong -\frac{i}{2\sqrt{2\pi}}\frac{\Delta\rho}{\sqrt{\rho_0}}\omega_k'[(d-x)\exp(i\omega_k'\sqrt{\rho_0}x) + x \cdot Sinc(\omega_k'\sqrt{\rho_0}x)] \ , \quad 0 \le x \le d. \quad (6.12)$$

the *Sinc*-function being defined as: $Sinc(\omega_k'\sqrt{\rho_0}x) = \sin(\omega_k'\sqrt{\rho_0}x)/\omega_k'\sqrt{\rho_0}x$.

*First-order Born approximated and actual QNM eigenfunctions.*

In this sub-section the following two properties are addressed:

- Even if the QNM eigenfunctions for a Fabry-Perot (FP) cavity satisfy eqs. (3.7) and (3.8), in any case the first-order Born approximation modifies the QNM eigenfunctions into approximated QNM eigenfunctions satisfying different boundary equations. This means that such an approximation eliminates the symmetry properties due to the presence of the term $(-1)^k$.

- The first-order Born approximation for an FP cavity introduces symmetry breaking: in fact, each Born approximated QNM eigenfunction produces values below or above the actual QNM eigenfunction value on the terminal surfaces of the same cavity. Consequently, the two error-functions for an approximated QNM are not equal in proximity to the two terminal surfaces of the cavity.

As from eq. (6.12), according to the first-order Born approximation for a Fabry-Perot (FP) cavity, the NM eigenfunctions and their first derivatives on the terminal surfaces of the cavity are calculated as:

$$\psi_k^{(nat)}(0) = -\frac{i}{2\sqrt{2\pi}} \frac{\Delta \rho}{\sqrt{\rho_0}} \omega_k' d , \qquad (6.13)$$

$$\psi_k^{(nat)}(d) = -\frac{i}{2\sqrt{2\pi}} \frac{\Delta \rho}{\sqrt{\rho_0}} \omega_k' d\, Sinc(\omega_k' \sqrt{\rho_0} d) , \qquad (6.14)$$

$$\partial_x \psi_k^{(nat)}(x)\Big|_{x=0} = \frac{1}{2\sqrt{2\pi}} (\omega_k')^2 \Delta \rho d , \qquad (6.15)$$

$$\partial_x \psi_k^{(nat)}(x)\Big|_{x=d} = -\frac{1}{2\sqrt{2\pi}} (\omega_k')^2 \Delta \rho d\, Sinc(\omega_k' \sqrt{\rho_0} d) . \qquad (6.16)$$

By combining the pairs of equations (6.13), (6.15) and (6.14), (6.16), the formal NM incoming waves conditions (4.8) and (4.9) are re-obtained, here reported for convenience:

$$\partial_x \psi_k^{(nat)}(x)\big|_{x=0} = i\omega_k' \sqrt{\rho_0} \psi_k^{(nat)}(0), \qquad (6.17)$$

$$\partial_x \psi_k^{(nat)}(x)\big|_{x=d} = -i\omega_k' \sqrt{\rho_0} \psi_k^{(nat)}(d). \qquad (6.18)$$

Thus, compared to the approximation (6.3) of the universe mode eigenfunctions, the first-order Born approximation (6.12) for a Fabry-Perot (FP) cavity better approximates the envelope of the NM eigenfunctions. In contrast with the universe mode approximation (6.3), the first-order Born approximation (6.12) satisfies the formal NM incoming waves conditions.

By combining the pairs of equations (6.13), (6.14) and (6.15), (6.16), according to the first-order Born approximation for an FP cavity, the link(s) between the pair(s) of values for the NM eigenfunctions (and for their derivatives) on the two terminal surfaces of the cavity are derived:

$$\psi_k^{(nat)}(d) = \psi_k^{(nat)}(0) Sinc(\omega_k' \sqrt{\rho_0} d), \qquad (6.19)$$

$$\partial_x \psi_k^{(nat)}(x)\big|_{x=d} = -\partial_x \psi_k^{(nat)}(x)\big|_{x=0} Sinc(\omega_k' \sqrt{\rho_0} d). \qquad (6.20)$$

Generally, for the physical interpretation of the fundamental theorem on NMs (see section 5.1), the NMs $[\omega_k', \psi_k^{(nat)}(x)]$ are correlated with the QNMs $[\omega_k, f_k(x) = f(x, \omega_k)]$ through the links $\omega_k' = -\omega_k$ and $\psi_k^{(nat)}(x) = f(x, -\omega_k')$; now, under the first-order Born approximation on the FP cavity [see eqs. (6.9) and (6.12)], even if the NM eigenfrequencies $\omega_k'$ are substituted with the QNM eigenfrequencies $(-\omega_k)$, the NM eigenfunctions $\psi_k^{(nat)}(x)$ are not converted into the actual QNM eigenfunctions $f_k(x)$ which satisfy eqs. (3.7) and (3.8), but into approximated QNM eigenfunctions $\tilde{f}_k(x)$ which satisfy modified relations:

$$\tilde{f}_k(d) = \tilde{f}_k(0) Sinc(\omega_k \sqrt{\rho_0} d), \qquad (6.21)$$

$$\partial_x \tilde{f}_k(x)\big|_{x=d} = -\partial_x \tilde{f}_k(x)\big|_{x=0} Sinc(\omega_k \sqrt{\rho_0} d). \qquad (6.22)$$

Even if the QNM eigenfunctions for an FP cavity satisfy eqs. (3.7) and (3.8), where the symmetry properties are evidenced by the presence of the term $(-1)^k$ in both the relations, in any case, under the first-order Born approximation, the approximated QNM eigenfunctions satisfy eqs. (6.21) and (6.22), where the symmetry properties have been lost due to the presence of the transfer function $Sinc(\omega_k \sqrt{\rho_0} d)$ in both the relations.

In the second of the eqs. (3.5), here reported for convenience,

$$\sin\left(\omega_k \sqrt{\rho} d\right) = -i(-1)^k \frac{2\sqrt{\rho \cdot \rho_0}}{\Delta \rho} ; \quad (6.23)$$

if the term $\rho$ is substituted by the term $\rho_0$, and so the term $\rho_0$ by the term $\rho$ and $\Delta \rho$ into $(-\Delta \rho)$, then the sinusoidal function can be calculated as

$$\sin\left(\omega_k \sqrt{\rho_0} d\right) = i(-1)^k \frac{2\sqrt{\rho \cdot \rho_0}}{\Delta \rho} , \quad (6.24)$$

and the "transfer function" for a Fabry-Perot (FP) cavity under the first-order Born approximation as:

$$Sinc\left(\omega_k \sqrt{\rho_0} d\right) = \frac{\sin\left(\omega_k \sqrt{\rho_0} d\right)}{\omega_k \sqrt{\rho_0} d} = (-1)^k 2i \frac{\sqrt{\rho}}{\Delta \rho} \frac{1}{\omega_k d}. \quad (6.25)$$

By inserting eq. (6.25), the eqs. (6.21) and (6.22) can be re-modelled as

$$\tilde{f}_k(d) = (-1)^k \tilde{f}_k(0) \cdot H_k(\rho, d), \quad (6.26)$$

$$\partial_x \tilde{f}_k(x)\big|_{x=d} = -(-1)^k \partial_x \tilde{f}_k(x)\big|_{x=0} \cdot H_k(\rho, d), \quad (6.27)$$

where $H_k(\rho, d)$ is a "weight-function" that can be considered as a correction for eqs. (3.7) and (3.8) due to the first-order Born approximation:

$$H_k(\rho, d) = 2i \frac{\sqrt{\rho}}{\Delta \rho} \frac{1}{\omega_k d}. \quad (6.28)$$

Applying the pair of equations (6.17) and (6.18), the eq. (6.20) is reduced in the eq. (6.19), so also the relations (6.21) and (6.22) are dependent and the eq. (6.21) is sufficient to calculate the "error-functions" for the $k^{th}$ QNM under the first-order Born approximation, next to the two terminal surfaces of a FP cavity:

$$\varepsilon_k(x=0) = \frac{\left|\tilde{f}_k(x=0)\right|^2 - \left|f_k(x=0)\right|^2}{\left|\tilde{f}_k(x=0)\right|^2} = 1 - \left|\frac{f_k(x=0)}{\tilde{f}_k(x=0)}\right|^2, \quad (6.29)$$

$$\varepsilon_k(x=d) = \frac{\left|\tilde{f}_k(x=d)\right|^2 - \left|f_k(x=d)\right|^2}{\left|\tilde{f}_k(x=d)\right|^2} = 1 - \left|\frac{f_k(x=d)}{\tilde{f}_k(x=d)}\right|^2. \quad (6.30)$$

Assuming the QNM eigenfunction on the left surface of the same cavity as an arbitrary constant, i.e.

$$f_k(x=0) = N', \quad (6.31)$$

and inserting $\omega'_n = -\omega_n$ into eq. (6.13) to obtain the approximated QNM value

$$\tilde{f}_k(0) = \frac{i}{2\sqrt{2\pi}} \frac{\Delta \rho}{\sqrt{\rho_0}} \omega_k d, \quad (6.32)$$

it follows that the error-function for the approximated $k^{th}$ QNM, in proximity to the left surface of a FP cavity is:

$$\varepsilon_k(x=0) = 1 - \frac{|N'|^2}{(\frac{1}{2\sqrt{2\pi}} \frac{\Delta \rho}{\sqrt{\rho_0}} |\omega_k| d)^2}. \quad (6.33)$$

Recalling eq. (3.7), the QNM eigenfunction on the right surface of the same cavity is

$$f_k(x=d) = (-1)^k N', \quad (6.34)$$

and applying eq. (6.26), the approximated QNM value is

$$\tilde{f}_k(d) = (-1)^k \frac{i}{2\sqrt{2\pi}} \frac{\Delta \rho}{\sqrt{\rho_0}} \omega_k d \cdot H_k(\rho, d), \quad (6.35)$$

following that the error-function for the approximated $k^{th}$ QNM, in proximity to the right surface of the FP cavity is:

$$\varepsilon_k(x=d) = 1 - \frac{|N'|^2}{(\frac{1}{2\sqrt{2\pi}} \frac{\Delta \rho}{\sqrt{\rho_0}} |\omega_k| d)^2} \cdot \frac{1}{|H_k(\rho, d)|^2}. \quad (6.36)$$

As from eqs. (6.33) and (6.36), the first-order Born approximation for an FP cavity introduces symmetry breaking: in fact, each Born approximated QNM eigenfunction produces values below or above the actual QNM eigenfunction value on the terminal surfaces of the same cavity. Consequently, the two error-functions for an approximated QNM are not equal in proximity to the two terminal surfaces of the cavity.

If the modulus of the arbitrary constant $N'$ is determined as

$$|N'| = \frac{i}{2\sqrt{2\pi}} \frac{\Delta\rho}{\sqrt{\rho_0}} |\omega_k| d, \quad (6.37),$$

then the error-function $\varepsilon_k(x=0)$ for the first-order Born approximated QNMs, in proximity to the left surface $x=0$ of the same cavity, is null

$$\varepsilon_k(x=0) = 0, \quad (6.38),$$

contrary to the error-function for the first-order Born approximated QNMs $\varepsilon_k(x=d)$, in proximity to the right surface $x=d$ of the same cavity, which is reduced to:

$$\varepsilon_k(x=d) = 1 - \frac{1}{|H_k(\rho,d)|^2} = 1 - \frac{1}{4}(\frac{\Delta\rho}{\sqrt{\rho}}d)^2(k^2\alpha^2 + \beta^2), \quad (6.39)$$

where α and β are given by eq. (3.2). As from eq. (6.39), the error-function for the approximated QNMs $\varepsilon_k(x=d)$ is a decreasing monotonous function when the QNM order-number $k^2$ is increasing, so the error-function assumes its maximum value in correspondence to the QNM of order-number $k=0$, i.e.

$$\varepsilon_0(x=d) = 1 - \frac{1}{4}(\frac{\Delta\rho}{\sqrt{\rho}}d)^2 \cdot \beta^2 = 1 - \frac{1}{4}(\frac{\Delta\rho}{\rho}\ln\frac{n+n_0}{n-n_0})^2, \quad (6.40)$$

and it is almost null in correspondence to the QNM the order-number of which is the integer-part of:

$$k = \pm\frac{1}{\alpha}\sqrt{4(\frac{\sqrt{\rho}}{\Delta\rho}\frac{1}{d})^2 - \beta^2} \cong \pm\frac{1}{\alpha}2(\frac{\sqrt{\rho}}{\Delta\rho}\frac{1}{d}) = \pm\frac{1}{\pi/2}\frac{1}{\Delta\rho/\rho}. \quad (6.41)$$

With reference to a Fabry-Perot (FP) cavity surrounded by the universe and characterized by a relative refractive index $n(\xi) = (1-\xi) \cdot n_0 + \xi \cdot n_0\sqrt{2}$, being $n_0$ the external refractive index and $\xi = 0.1$, sufficient to satisfy the condition of weak scattering [see eqs. (6.1) and (6.2)], the error-function for the QNMs under the first-order Born approximation, in proximity to the right surface of the FP cavity, assumes as maximum value $\varepsilon_0(x=d)=0.9769$ and it is almost null in correspondence to the QNM with order-number $k=8$ [fig. 2]; instead, the superposition, inside the whole cavity, of the square modulus $|f_k|^2$ for the actual QNMs, with order-number $k=0,1,2,\ldots$, on the corresponding approximated QNMs, reaches the optimal result in correspondence to the QNM with order-number $k'=k-1=7$ [fig. 3]: even if figs. 2 and 3 are affected by this minor shift about the optimal order-number, for the approximation of eq. (6.41), in any case the curves clearly evidence the error introduced by the first-order Born approximation on the envelopes of the QNM eigenfunctions.

## 7. Discussion and conclusive remarks.

In summary, this paper has proposed a comparison between the extinction theorem [1,2][8-11] and the Sturm-Liouville theory [14-20][21-24] approaches for calculating the electromagnetic (e.m.) field inside an optical cavity. We have discussed for the first time to the best of our knowledge, in the framework of classical electrodynamics, a simple link between the Quasi Normal Modes (QNMs) and the Natural Modes (NMs) for one-dimensional (1D), two-sided, open cavities. The main results obtained and discussed regard the field of application for the two bases. Two points seem worthy of note.

- The QNMs are defined for the problem of an electromagnetic (e.m.) field outgoing from an open cavity with a generic refractive index. It is natural to introduce the NMs only for the scattering problems of an e.m. field incoming within a medium that scatters rather weakly, i.e. when the refractive index of the medium differs only slightly from the value in the universe. As shown in section 4, the NM definition (4.6) satisfies the continuity conditions with suitable e.m. fields that exhibit an incoming wave behaviour near the cavity; in this approach, the cavity is interpreted as a scatterer. In addition, as discussed in section 5, starting from the physical interpretation for the incoming waves conditions [see eqs. (5.2) and (5.4)-(5.6)], it is plausible to assume the cavity as a medium which scatters very weakly, so that the refractive index of the cavity differs only slightly from the value in the universe.

- A further important general property of NMs is that they can exist in a medium without the presence of a driving field. This can be explained in terms of internal (electron) oscillations of the scatterer: the oscillations continue for some time even after the driving field is gone.

Finally, the QNM eigenfrequencies and eigenfunctions were calculated for a linear Fabry-Perot (FP) cavity. The first-order Born approximation was applied to the same cavity in order to compare between the first-order Born approximated and the actual QNM eigenfunctions of the cavity. We have demonstrated that the first-order Born approximation for an FP cavity introduces symmetry breaking: in fact, each Born approximated QNM eigenfunction produces values below or above the actual QNM eigenfunction value on the terminal surfaces of the same cavity. Consequently, the two error-functions for an approximated QNM are not equal in proximity to the two terminal surfaces of the cavity.

In conclusion, the novelty of this paper is the establishing of a link between QNM and NM bases and the consequent discussion about the application of the two expansion methods. In future, the choice between the two methods can be made starting from a coherent and solid theoretical base.

# Appendix A. The Green identities in the one dimensional (1D) universe.

With reference to fig. 1, an open cavity of length $d$ is now considered, filled with a material of refractive index $n(x)$ and enclosed in an infinite external space. The cavity also includes the terminal surfaces, so it is represented as $C = [0, d]$ and the rest of universe as $U = (-\infty, 0) \cup (d, \infty)$.

**First Green identity.**

It can easily be demonstrated that:

the Laplace transform $\tilde{E}(x, \omega)$ of the e.m. field, which solves eq. (2.1), and the Feynman propagator of the universe $G_0^{(F)}(x, x', \omega)$, which solves eq. (4.1) with the symmetry property (4.2) [27], satisfy the Green identities [30] in the one dimensional (1D) universe, the first inside the open cavity,

$$\tilde{E}(x, \omega) = \int_{0^-}^{d^+} \omega^2 [\rho(x') - \rho_0] \tilde{E}(x', \omega) G_0^{(F)}(x, x', \omega) dx' +$$

$$+ [\tilde{E}(x' = 0^-, \omega) \frac{\partial}{\partial x'} G_0^{(F)}(x, x', \omega) \bigg|_{x'=0^-} - G_0^{(F)}(x, x' = 0^-, \omega) \frac{\partial}{\partial x'} \tilde{E}(x', \omega) \bigg|_{x'=0^-} ] - \quad , \quad 0 \leq x \leq d,$$

$$- [\tilde{E}(x' = d^+, \omega) \frac{\partial}{\partial x'} G_0^{(F)}(x, x', \omega) \bigg|_{x'=d^+} - G_0^{(F)}(x, x' = d^+, \omega) \frac{\partial}{\partial x'} \tilde{E}(x', \omega) \bigg|_{x'=d^+} ]$$

(A.1)

and the second outside the same cavity,

$$\int_{0^-}^{d^+} \omega^2 [\rho(x') - \rho_0] \tilde{E}(x', \omega) G_0^{(F)}(x, x', \omega) dx' =$$

$$[\tilde{E}(x' = d^+, \omega) \frac{\partial}{\partial x'} G_0^{(F)}(x, x', \omega) \bigg|_{x'=d^+} - G_0^{(F)}(x, x' = d^+, \omega) \frac{\partial}{\partial x'} \tilde{E}(x, \omega) \bigg|_{x'=d^+} ] - \quad , \quad x < 0 \text{ or } x > d.$$

$$- [\tilde{E}(x' = 0^-, \omega) \frac{\partial}{\partial x'} G_0^{(F)}(x, x', \omega) \bigg|_{x'=0^-} - G_0^{(F)}(x, x' = 0^-, \omega) \frac{\partial}{\partial x'} \tilde{E}(x, \omega) \bigg|_{x'=0^-} ]$$

(A.2)

**Second Green identity.**

It can also be demonstrated that:

if the Laplace transform $\tilde{E}(x,\omega)$ for the e.m. field solves eq. (2.1) without specifying the boundary conditions, then the Feynman propagator of the universe $G_0^{(F)}(x,x',\omega)$ solves eq. (4.1) with the boundary conditions (4.4) and (4.5) [27], so the following identities are satisfied, the first inside the open cavity,

$$\tilde{E}(x,\omega) = \int_{0^-}^{d^+} \omega^2 [\rho(x') - \rho_0] \tilde{E}(x',\omega) G_0^{(F)}(x,x',\omega) dx' +$$
$$+ G_0^{(F)}(x,x'=0^-,\omega)[i\omega\sqrt{\rho_0}\tilde{E}(x'=0^-,\omega) - \frac{\partial}{\partial x'}\tilde{E}(x',\omega)\Big|_{x'=0^-}] + \quad , \text{ for } 0 \leq x \leq d, \quad (A.3)$$
$$+ G_0^{(F)}(x,x'=d^+,\omega)[i\omega\sqrt{\rho_0}\tilde{E}(x'=d^+,\omega) + \frac{\partial}{\partial x'}\tilde{E}(x',\omega)\Big|_{x'=d^+}]$$

the second on the left side of the same cavity,

$$\int_{0^-}^{d^+} \omega^2 [\rho(x') - \rho_0] \tilde{E}(x',\omega) G_0^{(F)}(x,x',\omega) dx' =$$
$$- G_0^{(F)}(x,x'=d^+,\omega)[i\omega\sqrt{\rho_0}\tilde{E}(x'=d^+,\omega) + \frac{\partial}{\partial x'}\tilde{E}(x',\omega)\Big|_{x'=d^+}] + \quad , \text{ for } x < 0, \quad (A.4)$$
$$+ G_0^{(F)}(x,x'=0^-,\omega)[i\omega\sqrt{\rho_0}\tilde{E}(x'=0^-,\omega) + \frac{\partial}{\partial x'}\tilde{E}(x',\omega)\Big|_{x'=0^-}]$$

and the third on the right side of the cavity,

$$\int_{0^-}^{d^+} \omega^2 [\rho(x') - \rho_0] \tilde{E}(x',\omega) G_0^{(F)}(x,x',\omega) dx' =$$
$$G_0^{(F)}(x,x'=d^+,\omega)[i\omega\sqrt{\rho_0}\tilde{E}(x'=d^+,\omega) - \frac{\partial}{\partial x'}\tilde{E}(x',\omega)\Big|_{x'=d^+}] - \quad , \text{ for } x > d. \quad (A.5)$$
$$- G_0^{(F)}(x,x'=0^-,\omega)[i\omega\sqrt{\rho_0}\tilde{E}(x'=0^-,\omega) - \frac{\partial}{\partial x'}\tilde{E}(x',\omega)\Big|_{x'=0^-}]$$

**Third identity.**

Finally, it can be demonstrated that:

if the Laplace transform $\tilde{E}(x,\omega)$ for the e.m. field solves eq. (2.1) without specifying the boundary conditions, then the following identities are satisfied:

$$\int_{0^-}^{d^+} \omega^2[\rho(x')-\rho_0]\tilde{E}(x',\omega)\exp(-i\omega\sqrt{\rho_0}x')dx' =$$

$$-\exp(-i\omega\sqrt{\rho_0}d)[i\omega\sqrt{\rho_0}\tilde{E}(x'=d^+,\omega)+\frac{\partial}{\partial x'}\tilde{E}(x',\omega)\bigg|_{x'=d^+}]+ \quad , \quad (A.6)$$

$$+[i\omega\sqrt{\rho_0}\tilde{E}(x'=0^-,\omega)+\frac{\partial}{\partial x'}\tilde{E}(x',\omega)\bigg|_{x'=0^-}]$$

$$\int_{0^-}^{d^+} \omega^2[\rho(x')-\rho_0]\tilde{E}(x',\omega)\exp(i\omega\sqrt{\rho_0}x')dx' =$$

$$\exp(i\omega\sqrt{\rho_0}d)[i\omega\sqrt{\rho_0}\tilde{E}(x'=d^+,\omega)-\frac{\partial}{\partial x'}\tilde{E}(x',\omega)\bigg|_{x'=d^+}]- \quad . \quad (A.7)$$

$$-[i\omega\sqrt{\rho_0}\tilde{E}(x'=0^-,\omega)-\frac{\partial}{\partial x'}\tilde{E}(x',\omega)\bigg|_{x'=0^-}]$$

## Appendix B.

**Proof of the first theorem on Natural Modes (NMs) [see eqs. (4.7)-(4.9)].**

From the definition (4.1) of the Green function over all the universe $G_0^{(F)}(x,x',\omega)$, the Natural Mode (NM) eigenfunctions solving the homogeneous integral equation (4.6) are the solutions of the differential equation

$$\left[\frac{d^2}{dx^2}+\rho_0(\omega'_n)^2\right]\psi_n^{(nat)}(x) = (\omega'_n)^2[\rho_0-\rho(x)]\psi_n^{(nat)}(x) \quad , \quad 0\leq x\leq d \quad , \quad (B.1)$$

which can be reduced to the differential Helmholtz equation (4.7) inside the open cavity.

Taking into account the Green identity (A.3) [see appendix A: second theorem], calculated in the eigenfrequencies $\omega = \omega'_n$ and for the eigenfunctions $\tilde{E}(x, \omega = \omega'_n) = \psi_n^{(nat)}(x)$ of the NMs, i.e.

$$\psi_n^{(nat)}(x) = \int_{0^-}^{d^+} (\omega'_n)^2 [\rho(x') - \rho_0] \psi_n^{(nat)}(x') G_0^{(F)}(x, x', \omega'_n) dx' +$$

$$+ G_0^{(F)}(x, x' = 0^-, \omega'_n)[i\omega'_n \sqrt{\rho_0} \psi_n^{(nat)}(x' = 0^-) - \frac{d}{dx'} \psi_n^{(nat)}(x') \Big|_{x'=0^-}] + \quad , \quad \text{for } 0 \le x \le d, \quad \text{(B.2)}$$

$$+ G_0^{(F)}(x, x' = d^+, \omega'_n)[i\omega'_n \sqrt{\rho_0} \psi_n^{(nat)}(x' = d^+) + \frac{d}{dx'} \psi_n^{(nat)}(x') \Big|_{x'=d^+}]$$

the necessary and sufficient condition to define the NM eigenfunctions as the eigensolutions of the homogeneous integral equation (4.6) is that the following equation

$$G_0^{(F)}(x, x' = 0^-, \omega'_n)[i\omega'_n \sqrt{\rho_0} \psi_n^{(nat)}(x' = 0^-) - \frac{d}{dx'} \psi_n^{(nat)}(x') \Big|_{x'=0^-}] +$$

$$+ G_0^{(F)}(x, x' = d^+, \omega'_n)[i\omega'_n \sqrt{\rho_0} \psi_n^{(nat)}(x' = d^+) + \frac{d}{dx'} \psi_n^{(nat)}(x') \Big|_{x'=d^+}] = 0 \quad \text{(B.3)}$$

is satisfied inside the cavity, i.e. $\forall x | 0 \le x \le d$. Eq. (B.3) holds inside the open cavity, if and only if the NMs satisfy the "incoming waves conditions" (4.8) and (4.9) next to the surfaces of the same cavity.

**Proof of the second theorem on NMs [see eqs. (4.10) and (4.11)].**

Taking into account the identity (A.6) [see appendix A: third theorem], calculated in correspondence to the eigenfrequencies $\omega = \omega'_n$ and for the eigenfunctions $\tilde{E}(x, \omega = \omega'_n) = \psi_n^{(nat)}(x)$ of the NMs, i.e.

$$\int_{0^-}^{d^+} (\omega'_n)^2 [\rho(x) - \rho_0] \psi_n^{(nat)}(x) \exp(-i\omega'_n \sqrt{\rho_0} x) dx =$$

$$- \exp(-i\omega'_n \sqrt{\rho_0} d)[i\omega'_n \sqrt{\rho_0} \psi_n^{(nat)}(x = d^+) + \frac{d}{dx} \psi_n^{(nat)}(x) \Big|_{x=d^+}] + \quad , \quad \text{(B.4)}$$

$$+ [i\omega'_n \sqrt{\rho_0} \psi_n^{(nat)}(x = 0^-) + \frac{d}{dx} \psi_n^{(nat)}(x) \Big|_{x=0^-}]$$

and by inserting in eq. (B.4) the NM incoming wave condition (4.9) next to the right surface of the open cavity, the value is deduced of the NM function (4.10) at this surface of the cavity. Analogously, taking into account the identity (A.7), calculated in the eigenfrequencies $\omega = \omega'_n$ and for the eigenfunctions $\tilde{E}(x, \omega = \omega'_n) = \psi_n^{(nat)}(x)$ of the NMs, i.e.

$$\int_{0^-}^{d^+} (\omega'_n)^2 [\rho(x) - \rho_0] \psi_n^{(nat)}(x) \exp(i\omega'_n \sqrt{\rho_0} x) dx =$$

$$\exp(i\omega'_n \sqrt{\rho_0} d) [i\omega'_n \sqrt{\rho_0} \psi_n^{(nat)}(x = d^+) - \frac{d}{dx} \psi_n^{(nat)}(x) \Big|_{x=d^+}] - \quad , \quad (B.5)$$

$$-[i\omega'_n \sqrt{\rho_0} \psi_n^{(nat)}(x = 0^-) - \frac{d}{dx} \psi_n^{(nat)}(x) \Big|_{x=0^-}]$$

and by inserting in eq. (B.5) the NM incoming wave condition (4.8) next to the left surface of the open cavity, the value is deduced of the NM function (4.11) at this surface of the cavity.

# Appendix C.

**Proof of the fundamental theorem on Natural Modes (NMs) [see eq. (5.1)].**

The Quasi Normal Modes (QNMs) of an open cavity are defined as pairs $[\omega_n, f_n(x)]$ with $n \in Z = \{0, \pm 1, \pm 2, ...\}$ such that eigenfunctions $f_n(x)$ solve the differential Helmholtz equation (2.4) inside the same cavity and satisfy the "formal" QNM "outgoing waves" equations (2.6) and (2.7) next to the terminal surfaces. Instead, for the first theorem on the Natural Modes (NMs) (see section 4.2.a), the NMs of an open cavity are defined as pairs $[\omega'_n, \psi_n^{(nat)}(x)]$ with $n \in Z = \{0, \pm 1, \pm 2, ...\}$ such that the eigenfunctions $\psi_n^{(nat)}(x)$ solve the same differential Helmholtz equation (4.7) inside the cavity but satisfy the "formal" NM "incoming waves" equations (4.8) and (4.9) next to the surfaces. The QNMs with eigenfrequencies $\omega_n$ can be correlated to the NMs with eigenfrequencies $\omega'_n$ through the link

$$\omega'_n = -\omega_n, \tag{C.1}$$

because the formal QNM outgoing waves equations (2.6) and (2.7) can be converted into the formal NM incoming waves equations (4.8) and (4.9), and vice-versa, simply applying eq. (C.1). The QNM eigenfrequencies $\omega_n$ are characterized by negative imaginary parts, $\mathrm{Im}\,\omega_n < 0$, and so the NM eigenfrequencies $\omega'_n$ are characterized by positive imaginary parts, $\mathrm{Im}\,\omega'_n > 0$.

**References.**

**Figures and captions**

Figure 1.

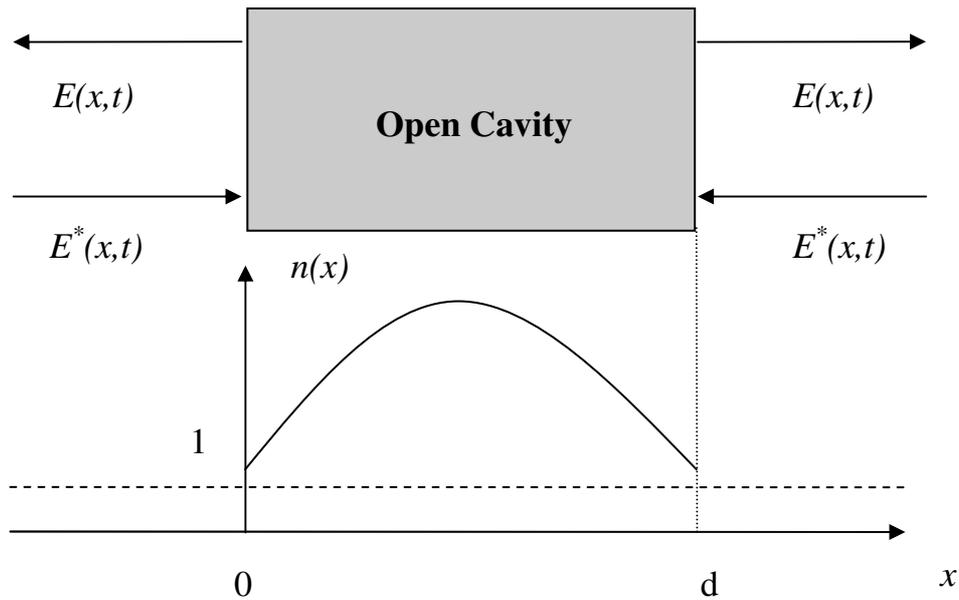

Figure 1. An open cavity of length $d$, filled with a refractive index $n(x)$ and enclosed in an infinite external space; the cavity also includes the terminal surfaces, so it is represented as $C=[0,d]$ and the rest of universe as $U=(-\infty,0)\cup(d,\infty)$. The two arrows labelled $E(x,t)$ represent the e.m. field satisfying the "outgoing waves" conditions (2.2) and (2.3); instead, the two arrows labelled $E^*(x,t)$ represent two external counter-propagating pumping fields, satisfying the "incoming waves conditions" (5.2) and (5.3).

Figure 2

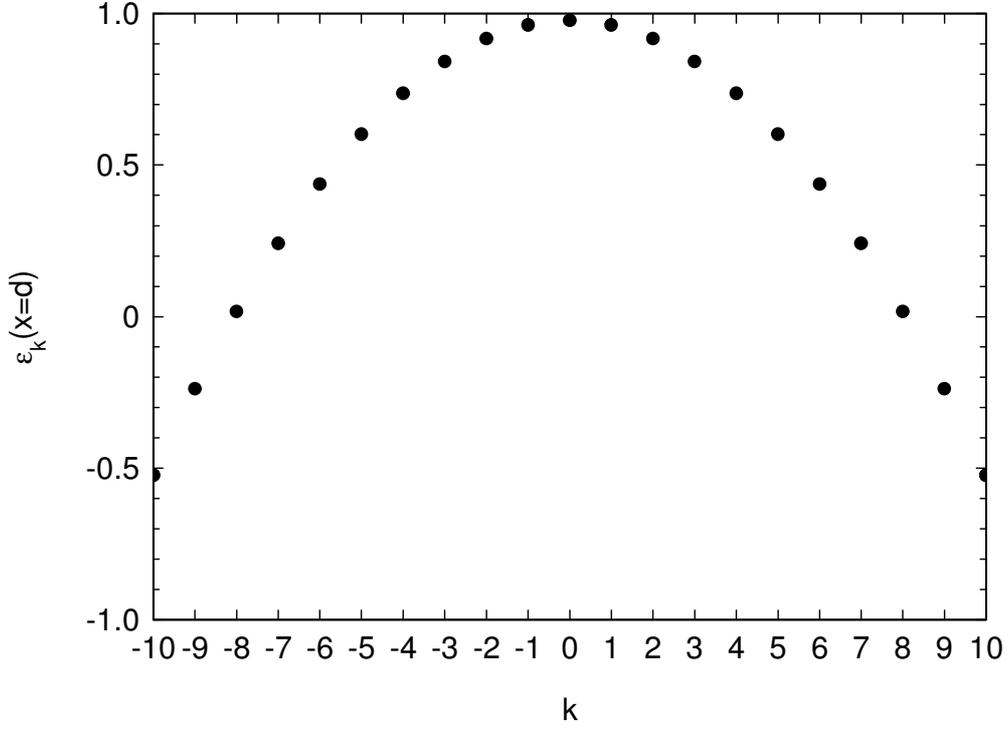

Figure 2. With reference to a Fabry-Perot (FP) cavity surrounded by the universe and characterized by a relative refractive index $n(\xi) = (1-\xi) \cdot n_0 + \xi \cdot n_0 \sqrt{2}$, being $n_0$ the external refractive index and $\xi = 0.1$, sufficient to satisfy the condition of weak scattering [see eqs. (6.1) and (6.2)]: plot for the "first-order Born" error $\varepsilon_k(x=d)$ as a function of the "approximated" QNM with order-number $k = 0, 1, \ldots, 10$, in proximity to the limiting right surface $x = d$ of the same cavity [see eq. (6.39)].

Figure 3.

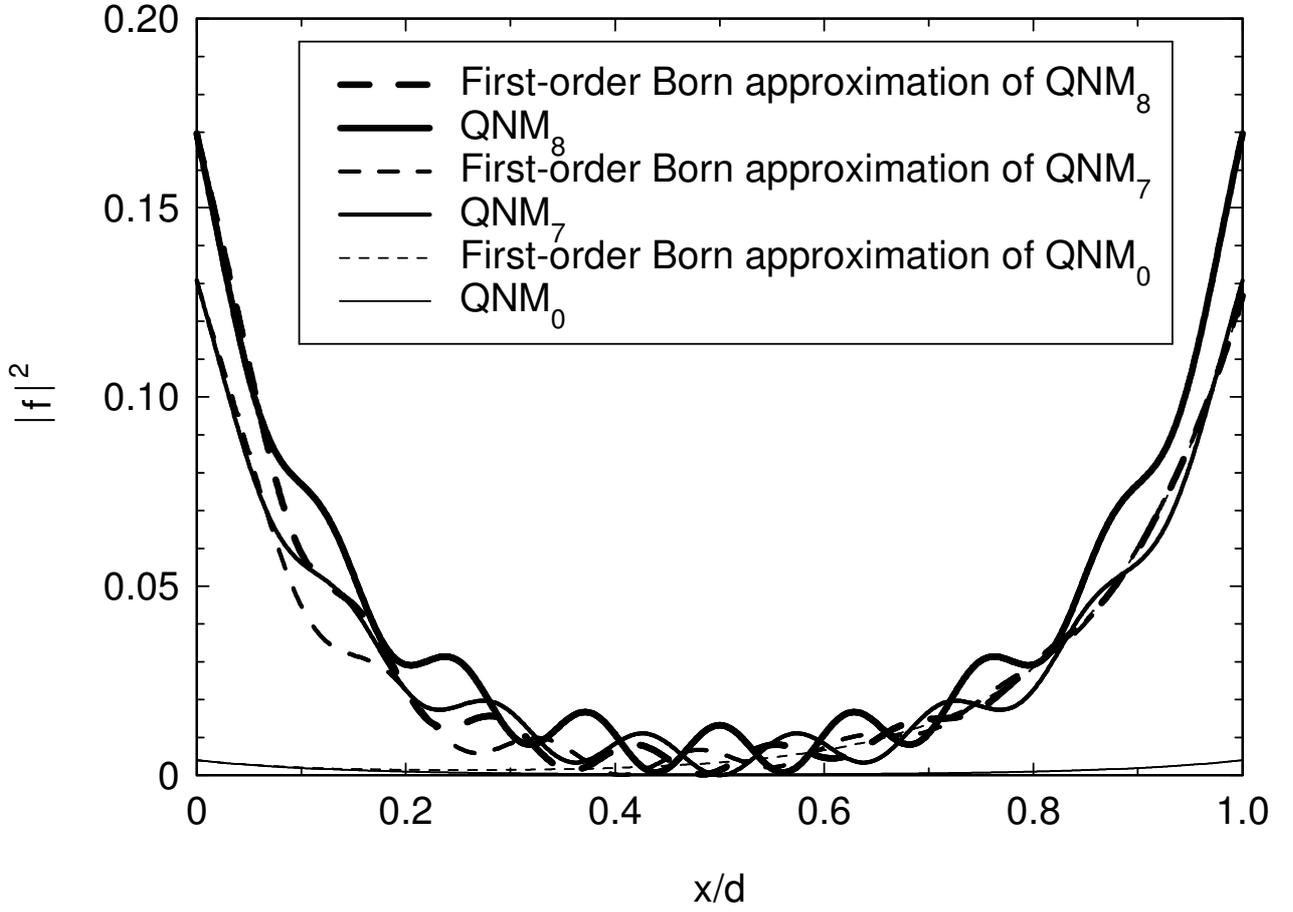

Figure 3. With reference to the FP cavity described in fig. 2: plots for the square modulus $\left|f_k\right|^2$ of the actual QNMs with order-number $k = 0, 7, 8$ [see eqs. (3.2)-(3.4)], superimposed on the corresponding modified QNMs [see eqs. (6.12) and (C.1)] after the first-order Born approximation [see eqs. (6.1) and (6.2)], as functions of the dimensionless space $x/d$, where $d$ is the length of the cavity.